\documentclass{article}

\usepackage{microtype}
\usepackage{graphicx}
\usepackage{subfigure}
\usepackage{booktabs} 
\graphicspath{{figs/}}
\usepackage{caption}
\usepackage{enumitem}

\usepackage{hyperref}

\usepackage[accepted]{icml2018}

\icmltitlerunning{Voice Imitating Text-to-Speech Neural Networks}

\begin{document}

\twocolumn[
\icmltitle{Voice Imitating Text-to-Speech Neural Networks}




\begin{icmlauthorlist}
\icmlauthor{Younggun Lee}{neo}
\icmlauthor{Taesu Kim}{neo}
\icmlauthor{Soo-Young Lee}{kaist}
\end{icmlauthorlist}

\icmlaffiliation{neo}{Neosapience, Seoul, Korea}
\icmlaffiliation{kaist}{Korea Advanced Institute of Science and Technology, Daejeon, Korea}

\icmlcorrespondingauthor{Younggun Lee}{yg@neosapience.com}
\icmlcorrespondingauthor{Taesu Kim}{taesu@neosapience.com}

\icmlkeywords{Machine Learning, Speech Synthesis, Text-to-speech, Voice imitation, Voice conversion}

\vskip 0.3in
]



\printAffiliationsAndNotice{}  

\begin{abstract}
We propose a neural text-to-speech (TTS) model that can imitate a new speaker's voice using only a small amount of speech sample. We demonstrate voice imitation using only a 6-seconds long speech sample without any other information such as transcripts. Our model also enables voice imitation instantly without additional training of the model. We implemented the voice imitating TTS model by combining a speaker embedder network with a state-of-the-art TTS model, Tacotron. The speaker embedder network takes a new speaker's speech sample and returns a speaker embedding. The speaker embedding with a target sentence are fed to Tacotron, and speech is generated with the new speaker's voice. We show that the speaker embeddings extracted by the speaker embedder network can represent the latent structure in different voices. The generated speech samples from our model have comparable voice quality to the ones from existing multi-speaker TTS models. \end{abstract}

\section{Introduction}
\label{sec_intro}
With recent improvements in deep neural network, researchers came up with neural network based vocoders such as WaveNet \cite{WAVENET1} and SampleRNN \cite{SAMPLERNN}. Those models showed their ability to generate high quality waveform from acoustic features. Some researchers further devised neural network based text-to-speech (TTS) models which can replace the entire TTS system with neural networks \cite{DV1, DV2, DV3, TACO1, TACO2}. Neural network based TTS models can be built without prior knowledge of a language when generating speech. Neural TTS models can be easily built compared to the previous approaches, which require carefully designed features, if we have enough (speech, text) pair data. Furthermore, the neural network based TTS models are capable of generating speech with different voices by conditioning on a speaker's index \cite{DV2, DV3} or an emotion label \cite{EMOTTS}.

Some researchers have tried to imitate a new speaker's voice using the speaker's recordings \cite{VOICELOOP}. Taigman et al. reported their model's ability to mimic a new speaker's voice by learning a speaker embedding of the new speaker \yrcite{VOICELOOP}. However, this approach requires additional training stage and transcriptions of the new speaker's speech sample. The transcription may not be always available, and the additional training stage prohibits immediate imitation of a new speaker's voice. In this study, we propose a voice imitating TTS model that can imitate a new speaker's voice without transcript of speech sample or additional training. This enables the voice imitation process immediately using only a short speech sample of a speaker. The proposed model takes two inputs: (1) target text and (2) a speaker's speech sample. The speech sample is first transformed into a speaker embedding by the speaker embedder network. Then a neural network based TTS model generates speech output by conditioning on the speaker embedding and the text input.

We implemented a baseline multi-speaker TTS model based on Tacotron, and we also implemented voice imitating TTS model by extending the baseline model. We investigated latent space of the learned speaker embeddings by visualizing with principal component analysis (PCA). We directly qualitatively compared similarity of voice from the both TTS models and the ground truth data. We further conducted two types of surveys to analyze the result quantitatively. The first survey compared generation quality of the voice imitating TTS and the multi-speaker TTS. The second survey checked how speaker-discriminable speech samples are generated by the both models. 

The main contributions of this study can be summarized as follows:
\vspace{-.5em}
\begin{enumerate}[leftmargin=2em]
	\item The proposed model makes it possible to imitate a new speaker's voice using only a 6-seconds long speech sample.
	\item Imitating a new speaker's voice can be done immediately without additional training.
	\item Our approach allows TTS model to utilize various sources of information by changing the input of the speaker embedder network.
\end{enumerate}

\section{Background}
\label{sec_related}

In this section, we review previous works that are related to our study. We will cover both traditional TTS systems and neural network based TTS systems. The neural network based TTS systems includes neural vocoder, single-speaker TTS, multi-speaker TTS, and voice imitation model.

\begin{figure*}[ht]
	\centerline{\includegraphics[width=14cm]{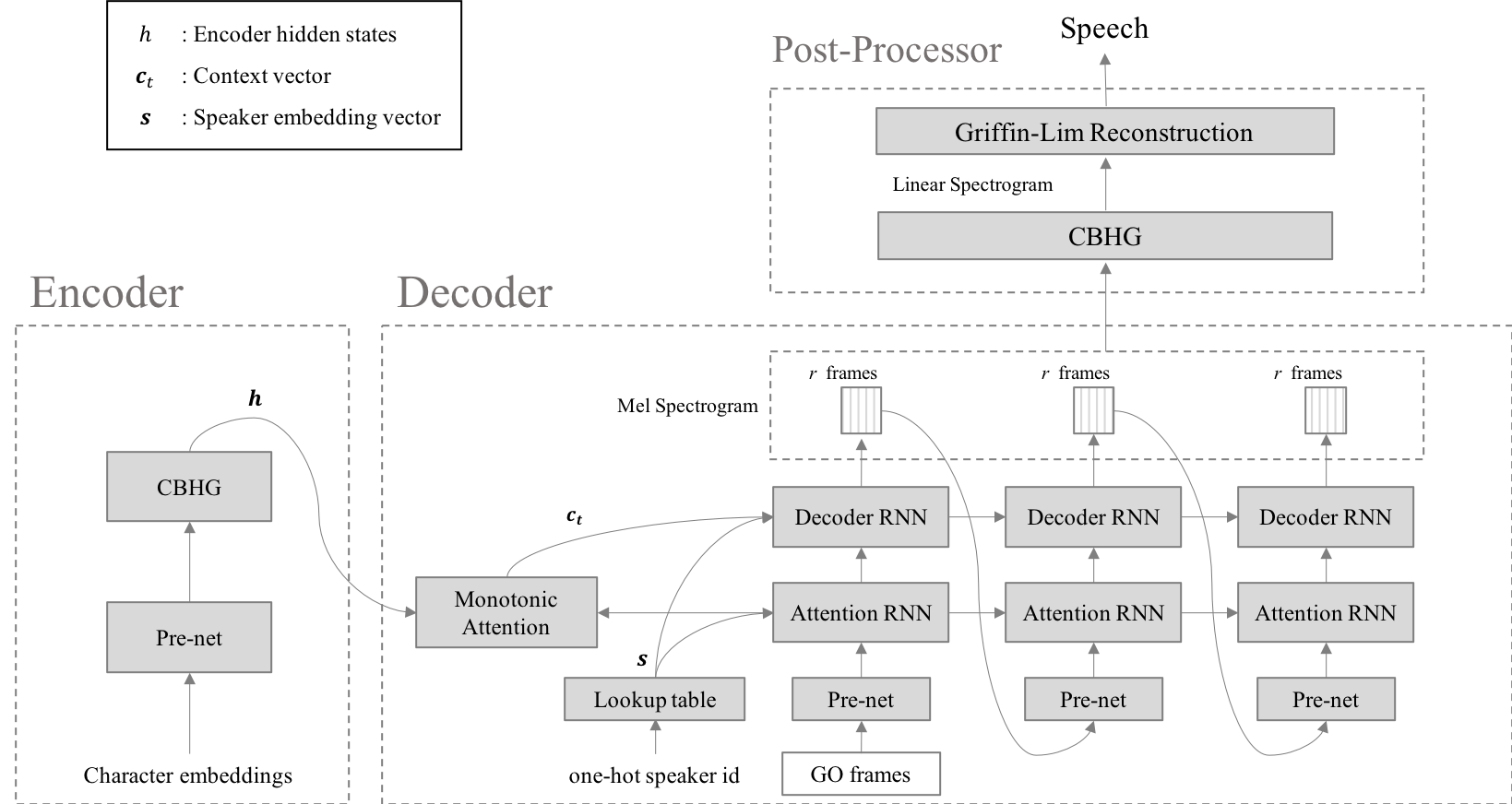}}
	\caption{Multi-speaker Tacotron}\label{fig-net1}
\end{figure*}

Common TTS systems are composed of two major parts: (1) text encoding part and (2) speech generation part. Using prior knowledge about the target language, domain experts have defined useful features of the target language and have extracted them from input texts. This process is called a text encoding part, and many natural language processing techniques have been used in this stage. For example, a grapheme-to-phoneme model is applied to input texts to obtain phoneme sequences, and a part-of-speech tagger is applied to obtain syntactic information. In this manner, the text encoding part takes a text input and returns various linguistic features. Then, the following speech generation part takes the linguistic features and generates waveform of the speech. Examples of the speech generation part include concatenative and parametric approach. The concatenative approach generates speech by connecting short units of speech which has a scale of phoneme or sub-phoneme level, and the parametric TTS utilizes a generative model to generate speech.

Having seen neural networks show great performance in regression and classification tasks, researchers have tried to substitute previously used components in TTS systems. Some group of researchers came up with neural network architectures that can substitute the vocoder of the speech generation part. Those works include Wavenet \cite{WAVENET1} and SampleRNN \cite{SAMPLERNN}. Wavenet can generate speech by conditioning on several linguistic features, and Sotelo et al. showed that SampleRNN can generate speech by conditioning on vocoder parameters \yrcite{CH2WAV}. Although these approaches can substitute some parts of the previously used speech synthesis frameworks, they still required external modules to extract the linguistic features or the vocoder parameters. Some researchers came up with neural network architectures that can substitute the whole speech synthesis framework. Deep Voice 1 \cite{DV1} is made of 5 modules where all modules are modelled using neural networks. The 5 modules exhaustively substitute the text encoding part and the speech generation part of the common speech synthesis framework. While Deep voice 1 is composed of only neural networks, it was not trained in end-to-end fashion.

Wang et al. proposed fully end-to-end speech synthesis model called Tacotron \yrcite{TACO1}. Tacotron can be regarded as a variant of a sequence-to-sequence network with attention mechanism \cite{SEQATT}. Tacotron is composed of three modules: encoder, decoder, and post-processor (refer to Figure \ref{fig-net1}). Tacotron basically follows the sequence-to-sequence framework with attention mechanism, especially which converts a character sequence into corresponding waveform. More specifically, the encoder takes the character sequence as an input and generates a text encoding sequence which has same length with the character sequence. The decoder generates Mel-scale spectrogram in an autoregressive manner. Combining attention alignment with the text encoding gives a context vector, and decoder RNN takes context vector and output of the attention RNN as inputs. The decoder RNN predicts Mel-scale spectrogram, and the post-processor module consequently generates linear-scale spectrogram from the Mel-scale spectrogram. Finally, Griffin-Lim reconstruction algorithm estimates waveform from the linear-scale spectrogram \cite{GLRECON}.

Single-speaker TTS systems have further extended to the multi-speaker TTS systems which can generate speech by conditioning on a speaker index. Arik et al. proposed Deep Voice 2, a modified version of Deep Voice 1, to enable multi-speaker TTS \yrcite{DV1, DV2}. By feeding learned speaker embedding as nonlinearity biases, recurrent neural network initial states, and multiplicative gating factors, they showed their model can generate multiple voices. They also showed Tacotron is able to generate multiple voice using the similar approach. Another study reported a TTS system that can generate voice containing emotions \cite{EMOTTS}. This approach is similar to the multi-speaker Tacotron in Deep Voice 2 paper, but the model could be built with less number of speaker embedding input connections. 

Multi-speaker TTS model is further extended to voice imitation model. Current multi-speaker TTS models takes a one-hot represented speaker index vector as an input, and this is not easily extendable to generate voices which are not in the training data. Because the model can learn embeddings only for the speakers represented by one-hot vectors, there is no way to get a new speaker's embedding. If we want to generate speech of a new speaker, we need to retrain the whole TTS model or fine-tune the embedding layer of the TTS model. However, training of the network requires large amount of annotated speech data, and it takes time to train the network until convergence. Taigman et al. proposed a model that can mimic a new speaker's voice \yrcite{VOICELOOP}. While freezing the model's parameters, they backpropagated errors using new speaker's (speech, text, speaker index) pairs to get a learned embedding. However, this model could not overcome the problems we mentioned earlier. The retraining step requires (speech, text) pair which can be inaccurate or even unavailable for data from the wild. Furthermore, because of the additional training, voice imitation cannot be done immediately. In this study, we will propose a TTS model that does not require annotated (speech, text) pairs so that it can be utilized in more general situations. Moreover, our model can immediately mimic a new speaker's voice without retraining.

\section{Voice imitating neural speech synthesizer}
\label{sec_method}

\subsection{Multi-speaker TTS}
\label{ssec_multiTTS}
One advantage to use neural network for a TTS model is that it is easy to give conditions when generating speech. For instance, we can give condition by just adding a speaker index input. Among several approaches to neural network based multi-speaker TTS models, we decided to adopt the architecture of Lee et al. \yrcite{EMOTTS}. Their model extends Tacotron to take a speaker embedding vector at the decoder of Tacotron (see Figure \ref{fig-net1}). If we drop the connections from the one-hot speaker ID input and the speaker embedding vector $s$, there is no difference from the original Tacotron architecture.  The model has two targets in its objective function: (1) Mel-scale spectrogram target $Y_{mel}$ and (2) linear-scale spectrogram target $Y_{linear}$. L1 distances of each Mel-scale spectrograms $\hat{Y}_{mel}$ and $Y_{mel}$ and linear-scale spectrograms $\hat{Y}_{linear}$ and $Y_{linear}$ are added to compute the objective function as follows:
\begin{equation}\label{eq_trainigobj}
Loss = ||Y_{linear}-\hat{Y}_{linear}||_1 + ||Y_{mel}-\hat{Y}_{mel}||_1
\end{equation}
where $\hat{Y}$'s are output of the Tacotron and $Y$'s are the ground truth spectrograms. Note that, there is no direct supervision on the speaker embedding vector, and each speaker index will have corresponding speaker embedding vector $s$ learned by backpropagated error from the loss function (\ref{eq_trainigobj}). By its formulation, the model can store only the speaker embeddings appeared in the training data at the Lookup table. When we want to generate a speech with a new speaker's voice, we need another speaker embedding for that speaker. In order to get a speaker embedding of the unseen speakers, we should train the model again with the new speaker's data. This retraining process consumes much time, and the model's usability limited to the voice with large data size. 

\begin{figure}[t]
	\centerline{\includegraphics[width=4cm]{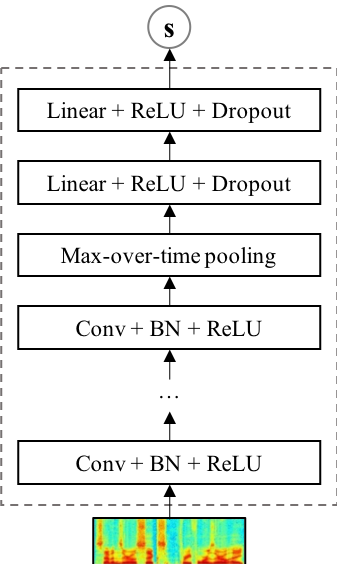}}
	\caption{The speaker embedder network}\label{fig-embednet}
\end{figure}

\begin{figure*}[t]
	\centerline{\includegraphics[width=14cm]{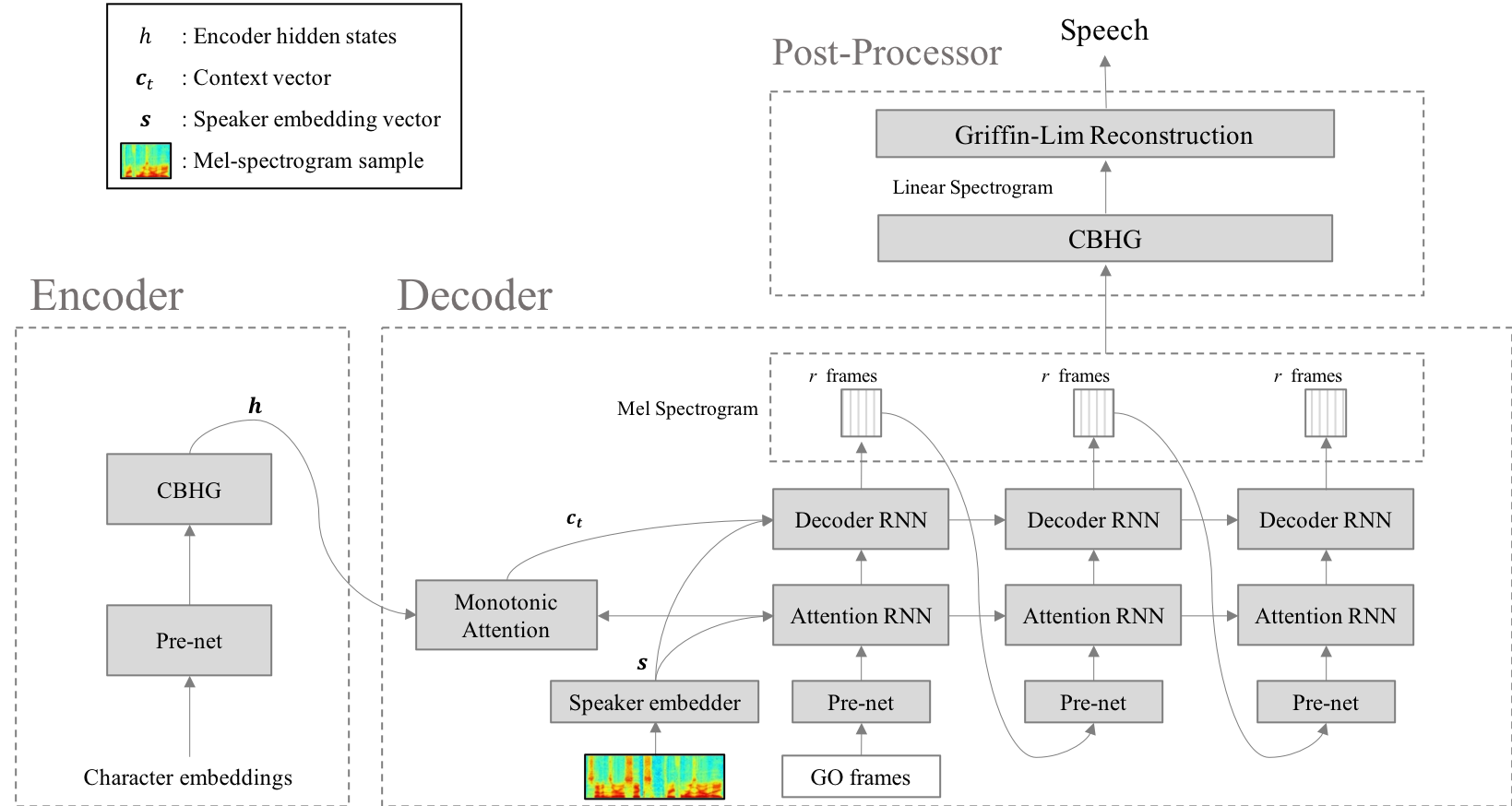}}
	\caption{Voice imitating Tacotron}\label{fig-net2}
\end{figure*}

\subsection{Proposed model}
\label{ssec_imiTTS}
One possible approach to address the problem is direct manipulation of the speaker embedding vector. Assuming the speaker embedding vector can represent arbitrary speakers' voices, we may get desired voice by changing values of the speaker embedding vector, but it will be hard to find the exact combination of values of the speaker embedding vector. This approach is not only inaccurate but also labor intensive. Another possible approach is to retrain the network using the new speaker's data. With enough amount of data, this approach can give us the desired speech output. However, it is not likely to have enough data of the new speaker, and the training process requires much time until convergence. To tackle this problem more efficiently, we propose a novel TTS architecture that can generate a new speaker’s voice using a small amount of speech sample. The imitation of a speaker's voice can be done immediately without requiring additional training or manual search of the speaker embedding vector.

The proposed voice imitating TTS model is an extension of the multi-speaker Tacotron in Section \ref{ssec_multiTTS}. We added a subnetwork that predicts a speaker embedding vector from a speech sample of a target speaker. Figure \ref{fig-embednet} shows the subnetwork, the speaker embedder, that contains convolutional layers followed by fully connected layers. This network takes log-Mel-spectrogram as input and predicts a fixed dimensional speaker embedding vector. Notice that, the input of speaker embedder network is not limited to the speech sample. Substituting input of the speaker embedder network enables TTS models to condition on various sources of information, but we focus on conditioning on a speaker's speech sample in this paper. 

Prediction of the speaker embedding vector requires only one forward pass of the speaker embedder network. This enables immediate imitation of the proposed model to generate speech for a new speaker. Although the input spectrograms may have various lengths, the max-over-time pooling layer, which is located at the end of the convolutional layers, squeezes the input into a fixed dimensional vector with length 1 for time axis. In this way, the voice imitating TTS model can deal with input spectrograms with arbitrary lengths. The speaker embedder with input speech sample replaces the Lookup table with one-hot speaker ID input of the multi-speaker Tacotron as described in Figure \ref{fig-net2}. For training of the voice imitating TTS model, we also use the same objective function (\ref{eq_trainigobj}) with the multi-speaker TTS. Note also that, there is no supervision on training the speaker embedding vector.

\section{Experiments}
\subsection{Dataset}
In accordance with Arik et al., we used VCTK corpus which contains 109 native English speakers with various accents \yrcite{DV2}. The population of speakers in VCTK corpus has various accents and ages, and each speaker recorded around 400 sentences.

We preprocessed the raw dataset in several ways. At first, we manually annotated transcripts for audio files which did not have corresponding transcripts. Then, for the text data, we filtered out symbols if they are not English letters or numbers or punctuation marks. We used capital letters without decapitalization. For the audio data, we trimmed silence using WebRTC Voice Activity Detector \cite{WEBRTC}. Reportedly, trimming silence is important for training Tacotron \cite{DV2,TACO1}. Note that, there is no label which tells the model when to start speech. If there is silence in the beginning of audio file, the model cannot learn what is the proper time to start speech. Removing silence can alleviate this problem by aligning the starting times of speeches. After the trimming, the total length of the dataset became 29.97 hours. Then, we calculated log-Mel-spectrogram and log-linear-spectrogram of each audio file. When generating spectrogram, we used Hann window of frame length 50ms and shifted windows by 12.5ms.

\subsection{Training}
In this experiment, we trained two TTS models: multi-speaker Tacotron and voice imitating Tacotron. In the rest of this paper, we will use terms multi-speaker TTS and voice imitating TTS to refer the two models respectively. To train the latter model, we did additional data preparation process. We prepared speech samples of each speaker since the model needs to predict a speaker embedding from log-Mel-spectrogram of a speech sample. Since we thought it is hard to capture a speaker’s characteristic in a short sentence, we concatenated one speaker’s whole speech data and made samples by applying fixed size rectangular window with overlap. The resulting window covers around 6 seconds of speech, which can contain several sentences. We fed a speech sample a target speaker to the model together with text input, while randomly drawing the speech sample from the windowed sample pool. We did not used the speech sample that is matched to the text input to prevent model from learning to generate by coping from the input speech sample. Furthermore, when training the voice imitating TTS model, we held out 10 speakers’ data for test set since we wanted to check if the model can generate unseen speakers’ voices. The profiles of 10 held out speakers are shown in Table \ref{tbl-testset}. We selected them to have similar distribution with training data in terms of gender, age, and accent.

\begin{table}[t]
	\caption{Speaker profiles of the test set.}
	\label{tbl-testset}
	\vskip 0.15in
	\begin{center}
		\begin{small}
			\begin{sc}
				\begin{tabular}{ccccc}
					\toprule
					ID & Age & Gender & Accents & Region \\
					\midrule
					225 & 23 & F &  English  & S. England \\
					226 & 22 & M &  English  & Surrey \\
					243 & 22 & M &  English  & London \\
					244 & 22 & F &  English  & Manchester\\
					262 & 23 & F &  Scottish & Edinburgh \\
					263 & 22 & M &  Scottish & Aberdeen \\
					302 & 20 & M &  Canadian & Montreal \\
					303 & 24 & F &  Canadian & Toronto \\
					360 & 19 & M &  American & New Jersey \\
					361 & 19 & F &  American & New Jersey \\
					\bottomrule
				\end{tabular}
			\end{sc}
		\end{small}
	\end{center}
	\vskip -0.1in
\end{table}

For the Tacotron’s parameters, we basically followed specifications written in the original Tacotron paper except the reduction factor $r$ \cite{TACO1}. We used 5 for the $r$, which means generating 5 spectrogram frames at each time-step. For hyperparameters of the speaker embedder network, we used the following settings. We used 5-layered 1D-convolutional neural network with 128 channels and window size of 3. The first 2 layers have stride of 1 and the remaining 3 layers have stride of 2. We used 2 linear layers with 128 hidden units after the convolutional layers. We used ReLU as a nonlinearity and applied batch normalization for every layer \cite{BN}. We also applied dropout with ratio of 0.5 to improve generalization \cite{DROPOUT}. The last layer of the speaker embedder network is a learnable projection layer without nonlinearity and dropout.

We used mini-batch size of 32. During the training, limited capacity of GPU’s memory prevented us from loading a mini-batch of long sequences at once. To maximize the utilization of data, we used truncated backpropagation through time \cite{TBPTT}. We used gradient clipping to avoid the exploding gradient problem \cite{CLIPPING}. We used 1.0 as a clipping threshold. For the optimization, we used ADAM \cite{ADAM}, which adaptively changes scale of update, with parameters 0.001, 0.9, and 0.999 for learning rate, $\beta_1$, and $\beta_2$ respectively.

\section{Result}

\begin{table}[t]
	\begin{center}
		\begin{tabular} {cc}
			\parbox[c]{4cm}{\includegraphics[width=4cm]{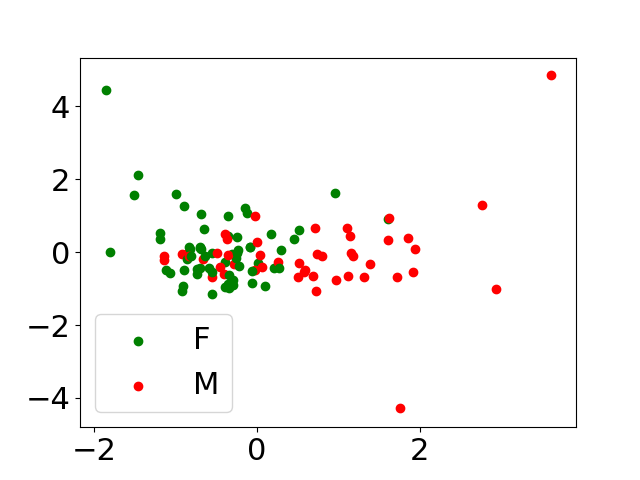}}&
			\parbox[c]{4cm}{\includegraphics[width=4cm]{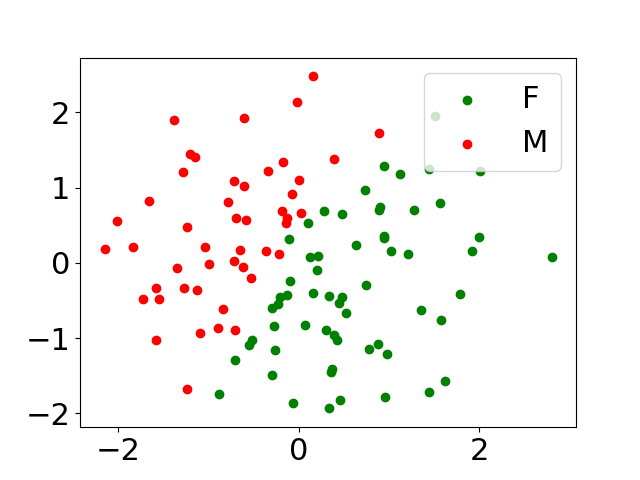}}\\
			(a) Voice imitating TTS & (b) Multi-speaker TTS\\
		\end{tabular}
	\end{center}
	\captionof{figure}	{Principal component of speaker embeddings of voice imitating TTS and multi-speaker TTS, shown with gender of the speakers}
	\label{fig_pca}
\end{table}

\begin{table*}[t]
\begin{center}
	\begin{tabular} {ccc}
		\parbox[c]{5.5cm}{\includegraphics[width=5.5cm]{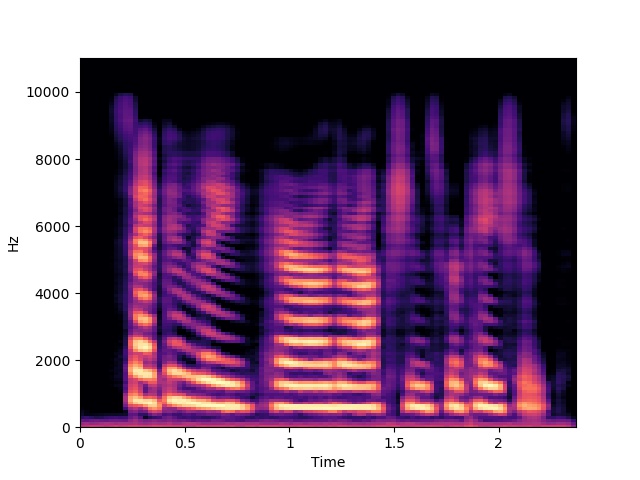}}&
		\parbox[c]{5.5cm}{\includegraphics[width=5.5cm]{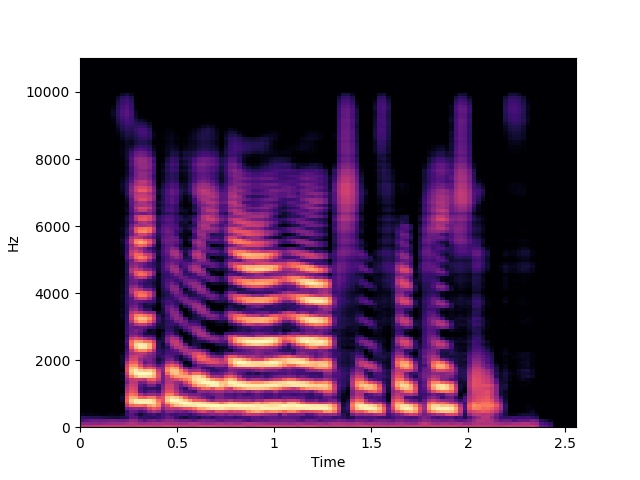}}&
		\parbox[c]{5.5cm}{\includegraphics[width=5.5cm]{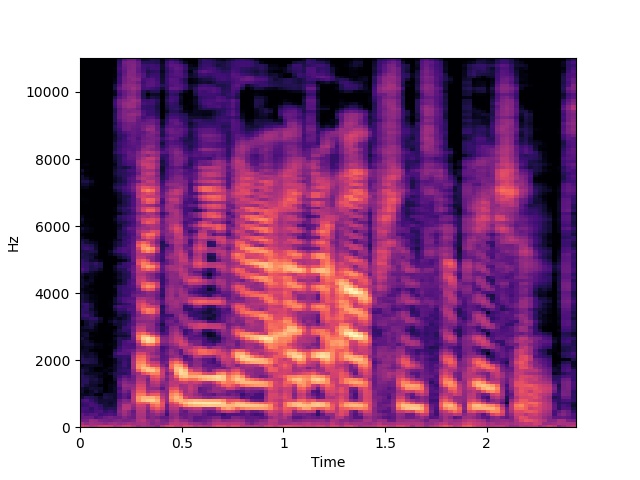}}\\
		Voice imitating TTS & Multi-speaker TTS & Ground truth \\
	\end{tabular}
\end{center}
\captionof{figure}	{Mel-spectrogram from Multi-speaker TTS and voice imitating TTS, generated for train set}
\label{fig_mimic1}
\end{table*}

\begin{table*}[t]
\begin{center}
	\begin{tabular} {ccc}
		\parbox[c]{5.5cm}{\includegraphics[width=5.5cm]{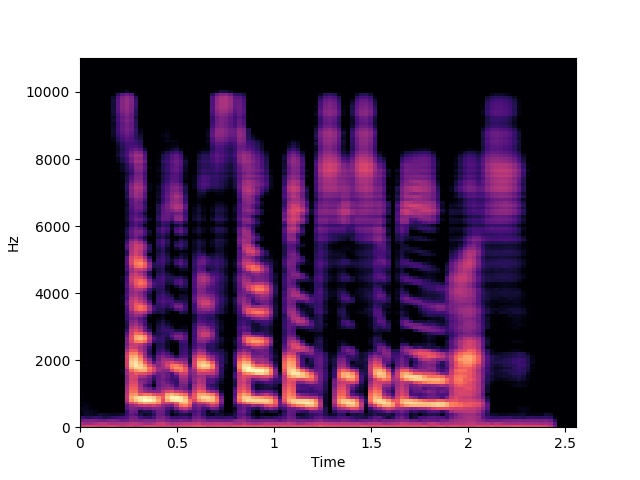}}&
		\parbox[c]{5.5cm}{\includegraphics[width=5.5cm]{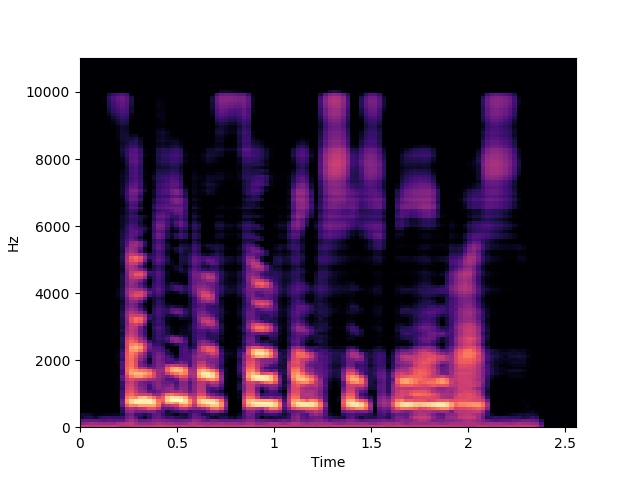}}&
		\parbox[c]{5.5cm}{\includegraphics[width=5.5cm]{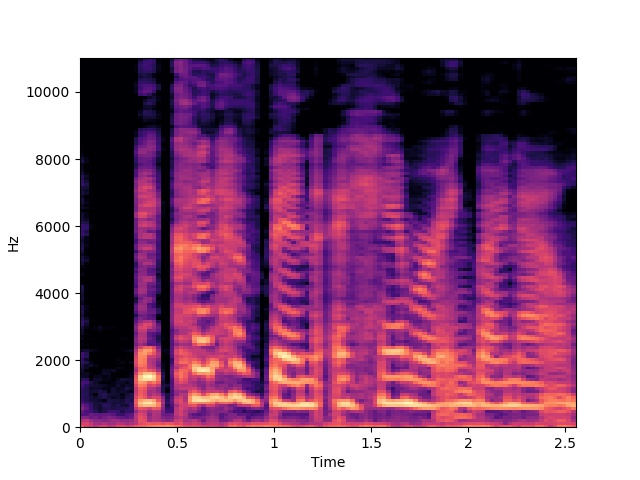}}\\
		Voice imitating TTS & Multi-speaker TTS & Ground truth\\
	\end{tabular}
\end{center}
\captionof{figure}	{Mel-spectrogram from Multi-speaker TTS and voice imitating TTS, generated for test set}
\label{fig_mimic2}
\end{table*}

We first checked performance of voice imitating TTS qualitatively by investigating learned latent space of the speaker embeddings. In order to check how the speaker embeddings are trained, we applied PCA to the speaker embeddings. Previous researches reported discriminative patterns were found from the speaker embeddings in terms of gender and other aspects \cite{DV2, VOICELOOP}. Figure \ref{fig_pca} shows the first two principal components of the speaker embeddings where green and red colors represent female and male respectively. We could see clear separation from speaker embeddings of the multi-speaker TTS as reported from other studies. Although the speaker embeddings of voice imitating TTS had an overlapped area, we could observe that the female embeddings are dominant in the left part, whereas the male embeddings are dominant in the right part. Besides, some of the embeddings located far from the center. We suspect that the overlap and the outliers are existing because the speaker embedding is extracted from a randomly chosen speech sample of a speaker. A speech sample of a male speaker can have only the particularly lower-pitched voice, or a speech sample of a female speaker can have only particularly higher-pitched voice. This may result in prediction of the out-lying embeddings, and similar argument could be applied for the overlapping embeddings.

To check how similar are the generated voices and the ground truth voice, we compared spectrogram and speech samples from the voice imitating TTS to that of multi-speaker TTS model and the ground truth data. Then, by feeding a text from the training data while conditioning on the same speaker, we generated samples from voice imitating TTS and multi-speaker TTS. Then we compared the generated samples and the corresponding ground truth speech samples. Example spectrograms from the both models and the ground truth data are shown in Figure \ref{fig_mimic1}. We could observe both models gave us similar spectrogram, and also the difference between them was negligible when we listened to the speech samples. From the spectrogram, we could observe they have similar pitch and speed by seeing heights and widths of harmonic patterns. When we compared generated samples of the both models to the ground truth data, we could observe the samples from the both models having simliar pitch with the ground truth. We could see the model can learn to predict speaker embedding from the speech samples.

Similarly, we analyzed spectrograms to check whether the voice imitating TTS can generalize well on the test set. Note that, the multi-speaker TTS included the test set of the voice imitating TTS for its training data, because otherwise multi-speaker TTS cannot generate speech for unseen speakers. In Figure \ref{fig_mimic2}, we also could observe spectrograms from generated samples showing similar pattern, especially for the pitch of each speaker. With these results, we conjecture that the model at least learned to encode pitch information in the speaker embedding, and it was generalizable to the unseen speakers. 

\begin{figure}[t]
	\centerline{\includegraphics[width=7cm]{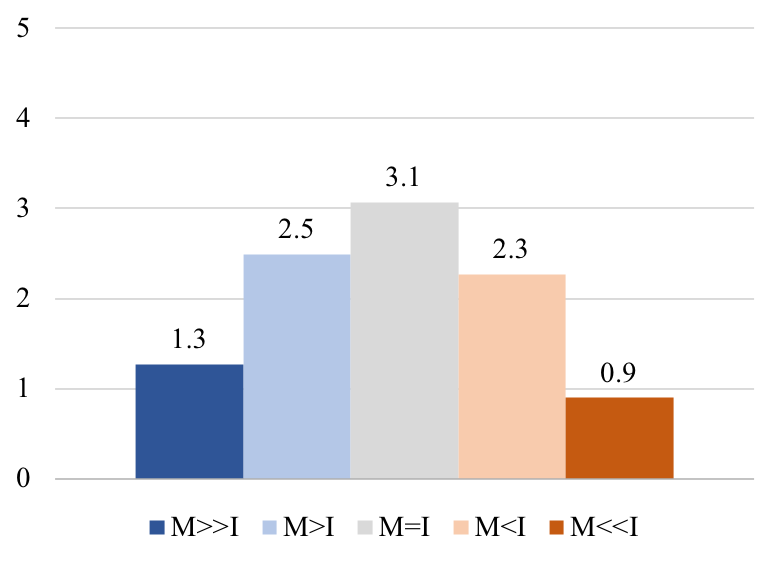}}
	\caption{Average scores of generated sample comparison survey, where the symbols mean: M$>>$I - M (the multi-speaker TTS) is far better than I (the voice imitating TTS), M$>$I - M is little better than I, M$=$I - Both M and I have same quality }\label{fig-survey1}
\end{figure}

Since it is difficult to evaluate generated speech sample objectively, we conducted surveys using crowdsourcing platforms such as Amazon's Mechanical Turk. We first made speech sample comparison questions to evaluate voice quality of generated samples. This survey is composed of 10 questions. For each question, 2 audio samples--one from the voice imitating TTS and the other one from the multi-speaker TTS--are presented to participants, and the participants are asked to give a score from -2 (multi-speaker TTS is far better than voice imitating TTS) to 2 (multi-speaker TTS is far worse than voice imitating TTS). We gathered 590 ratings on the 10 questions from 59 participants (see Figure \ref{fig-survey1}). From the result, we could observe the ratings were concentrated on the center with overall mean score of $-0.10 \pm 1.16$. It seems there is not much difference in the voice quality of the voice imitating TTS and the multi-speaker TTS. 

For the second survey, we made speaker identification questions to check whether generated speech samples contain distinct characteristics. The survey consists of 40 questions, where each question has 3 audio samples: ground truth sample and two generated samples. The two generated samples were from the same TTS model, but each of which conditioned on different speakers' index or speech samples. The participants are asked to choose one speech sample that sounds mimicking the same speaker identity of the ground truth speech sample. From the crowdsourcing platform, we found 50 participants for surveying the voice imitating TTS and other 50 participants for surveying the multi-speaker TTS model. The resulted speaker identification accuracies were 60.1$\%$ and 70.5$\%$ for the voice imitating TTS and the multi-speaker TTS respectively. Considering random selection will score 50$\%$ of accuracy, we may argue higher accuracies than 50$\%$ reflect distinguishable speaker identity in the generated speech samples. By its nature of the problem, it is more difficult to generate distinct voice for the voice imitating TTS. Because the voice imitating TTS must capture a speaker's characteristic in a short sample whereas the multi-speaker TTS can learn the characteristic from vast amount of speech data. Considering these difficulties, we think the score gap between the two models are explainable.

\section{Conclusion}
We have proposed a novel architecture that can imitate a new speaker's voice. In contrast to the current multi-speaker speech synthesis models the voice imitating TTS could generate a new speaker's voice using a small amount of speech sample. Furthermore, our method could imitate voice immediately without additional training. We have evaluated generation performance of the proposed model both in qualitatively and quantitatively, and we have found there is no significant difference in the voice quality between the voice imitating TTS and the multi-speaker TTS. Though generated speech from the voice imitating TTS have showed less distinguishable speaker identity than that from the multi-speaker TTS, generated voices from the voice imitating TTS contained pitch information which can make voice distinguishable from other speakers' voice. 

Our approach is particularly differentiated from the previous approaches by learning to extract features with the speaker embedder network. Feeding various sources of information to the speaker embedder network makes TTS models more versatile, and exploring its possibility is connected to our future works. We expect intriguing researches can be done in the future by extending our approach. One possible direction will be a multi-modal conditioned text-to-speech. Although this paper has focused on extracting speaker embedding from a speech sample, the speaker embedder network can learn to extract speaker embedding from various sources such as video. In this paper, the speaker embedder network has extracted a speaker's characteristic from a speech sample. By applying same approach to a facial video sample, the speaker embedder network may capture emotion or other characteristics from the video sample. The resulting TTS system will be able to generate a speaker's voice which contains appropriate emotion or characteristics for a given facial video clip and an input text. Another direction will be cross-lingual voice imitation. Since our model requires no transcript corresponding to the new speaker's speech sample, the model has a potential to be applied in the cross-lingual environment. For instance, imitating a Chinese speaker's voice to generate English sentence can be done. 

%
%


\bibliography{imitativeTTS}
\bibliographystyle{icml2018}
\end{document}